\documentclass{myjournal}%[a4paper,UKenglish]{lipics-v2016}
\usepackage{microtype}%if unwanted, comment out or use option "draft"
\newcommand{\frage}[1]{{\sf[ #1]}\marginpar{?}}
\newcommand{\psfrage}[1]{\frage{ps:#1}}
\renewcommand{\frage}[1]{}

\newcommand{\myparagraph}[1]{\subparagraph*{#1}}%\noindent{\bf #1}}
\usepackage{xcolor,hyperref}
\usepackage{graphicx}

\title{Dynamic Space Efficient Hashing} % -- a novel approach to growing!}
\author{Tobias Maier and Peter Sanders}
\institute{Karlsruhe Institute of Technology, Karlsruhe, Germany \quad
                \email{\{t.maier,sanders\}@kit.edu}}

\usepackage{amsmath}
\usepackage{amssymb}

\begin{document}

\maketitle

\begin{abstract}
We consider space
efficient hash tables that can grow and shrink dynamically and are
always highly space efficient, i.e., their space consumption is always
close to the lower bound even while growing and when taking into
account storage that is only needed temporarily. None of the
traditionally used hash tables have this property. We show how known
approaches like linear probing and bucket cuckoo hashing can be
adapted to this scenario by subdividing them into many subtables or
using virtual memory overcommitting. However, these rather
straightforward solutions suffer from slow amortized insertion times
due to frequent reallocation in small increments.

Our main result is DySECT ({\bf Dy}namic {\bf S}pace {\bf E}fficient
{\bf C}uckoo {\bf T}able) which avoids these problems. DySECT consists
of many subtables which grow by doubling their size. The resulting
inhomogeneity in subtable sizes is equalized by the flexibility
available in bucket cuckoo hashing where each element can go to
several buckets each of which containing several cells. Experiments
indicate that DySECT works well with load factors up to 98\%.  With
up to 2.7 times better performance than the next best solution.

%% Hash table research has concentrated on optimizing speed and space
%% efficiency for a long time.  Modern open addressing
%% techniques allow hash tables to be filled to more than 95\% before performance
%% deterioration becomes a significant issue.  Once this threshold is met, copying and
%% migrating the entire table still seems to be the most efficient answer.
%% Growing by a constant factor $\gamma$ often leads to bad space
%% efficiency because the filling degree of the grown table is at most
%% $\gamma^{-1}$. During the migration space efficiency is even
%% worse, since oftentimes both the new and the old table coexist at the
%% same time, causing a massive space overhead.

%% We propose a new hash table architecture that supports growing while
%% guaranteeing a strict size constraint $\alpha \cdot n$ throughout its
%% lifetime.  Nonetheless, we achieve guaranteed constant lookup times
%% and amortized constant insertion times.  Our architecture splits the
%% overall hash table into subtables that grow independently of each
%% other.  The cuckoo load balancing mechanism can move elements between
%% subtables, ensuring that growing one subtable increases the overall
%% insertion opportunities.
 \end{abstract}

\section{Introduction}
Dictionaries represented as hash tables are among the most frequently
used data structures and often play a critical role in achieving high
performance.  Having several compatible implementations, which perform
well under different conditions and can be interchanged freely, allows
programmers to easily adapt known solutions to new circumstances.

One aspect that has been subject to much investigation is space
efficiency~\cite{CuckooHashingFurtherAnalysis,
  TightThresholdsForCuckooHashingViaXorSat,
  BalancedAllocationAndDictionariesWithTightlyPackedConstantSizeBin,
  SpaceEfficientWithWorstCaseConstantAccess,
  SomeOpenQuestions, CuckooHashing}.
Modern space efficient hash tables work well even when filled to 95\%
and more.  To reach filling degrees like this, the table has to be
initialized with the correct final capacity, thereby, requiring
programmers to know tight bounds on the maximum number of inserted
elements.  This is typically not realistic.  For example, a frequent application of hash tables
aggregates information about data elements by their key.  Whenever the
exact number of unique keys is not known a priori, we have to
overestimate the initial capacity to guarantee good performance.
Dynamic space efficient data structures are necessary to guarantee
both good performance and low overhead independent of the
circumstances.

To visualize this, assume the following scenario.  During a word count
benchmark, we know an upper bound $n_{\max}$ to the number of unique
words. Therefore, we construct a hash table with at least $n_{\max}$
cells. If an instance only contains $0.7\cdot n_{max}$ unique words,
no static hash table can fill ratios greater than 70\%. Thus, dynamic
space efficient hash tables are required to achieve guaranteed
near-optimal memory usage.
%To achieve space efficiency
In scenarios where the final size is not known, the hash table has to
grow closely with the actual number of elements.
%\psfrage{end shortening potential}
This cannot be achieved efficiently with any of the current techniques
used for hashing and migration.

Many libraries -- even ones that implement space efficient hash tables
-- offer some kind of growing mechanism.  However, all existing
implementations either lose their space efficiency or suffer from
degraded performance once the table grows above its original capacity.
Growing is commonly implemented either by creating additional hash
tables -- decreasing performance especially for lookups or by
migrating all elements to a new table -- losing the space efficiency
by multiplying the original size.

To avoid the memory overhead of full table migrations, during which
both the new and the old table coexist, we propose an in-place growing
technique that can be adapted to most existing hashing schemes.
However, frequent migrations with small relative size changes remain
necessary to stay space efficient at all times.

To avoid both of these pitfalls we propose a variant of (multi-way)
bucket cuckoo
hashing~\cite{BalancedAllocationAndDictionariesWithTightlyPackedConstantSizeBin,
  SpaceEfficientWithWorstCaseConstantAccess}.  A technique where each
element can be stored in one of several associated constant sized
buckets.  When all of them are full, we move an element into one of
its other buckets to make space.  To solve the problem of efficient
migration, we split the table into multiple subtables, each of which
can grow independently of all others.  Because the buckets associated
with one element are spread over the different subtables, growing one
subtable alleviates pressure from all others by allowing moves from a
dense subtable to the newly-grown subtable.

Doubling the size of one subtable increases the overall size only by a
small factor while moving only a small number of elements. This makes
the size changes easy to amortize. The size and occupancy imbalance
between subtables (introduced by one subtable growing) is alleviated
using displacement techniques common to cuckoo hashing. This allows
our table to work efficiently at fill rates exceeding 95\%.

\psfrage{todo: sell soft real time}

%\psfrage{todo: sell subtable and VM overcommitting}

\psfrage{todo: paper overview}

%% \section{Motivation}
%% \label{sec:motivation}
%% Space efficiency is a major factor for the development of new hashing
%% techniques.  The goal has long been to make hash tables work better in
%% high load scenarios.  Modern hashing techniques like Hopscotch, Robin
%% Hood, and Cuckoo Hashing can easily fill hash tables to more than 90\%
%% and up to 98\% while still achieving constant average running times on
%% most of their operations.  But since these techniques depend on
%% statically sized tables this only makes sense, if the number of unique
%% elements can be estimated with similarly tight bounds.

%% To visualize this assume the following scenario.  During the operation
%% of a word count benchmark, we know an upper bound $c$ to the number of
%% unique words. Therefore, we construct a table large enough to hold
%% these elements ($0.9^{-1}\cdot c$ cells).  It is entirely possible
%% that the specific word count instance only contains $0.7\cdot c$
%% unique words (or less).  The resulting table is therefore only filled
%% 63\%.  It is clear that better static sized tables cannot achieve fill
%% ratios greater than 70\%.

%% Depending on the underlying problem it is impossible to tightly
%% estimate the final number of elements. Since dynamic growing is
%% independent from the problem's domain, it can be a solution even in
%% these instances.  To achieve space efficiency in scenarios where the
%% final size is not known the hash table has to grow closely with the
%% actual number of elements.  This is not efficient with any of the
%% current techniques used for hashing and migration.

We begin our paper by presenting some previous work (Section~\ref{sec:rel}).  Then we go into some notations
(Section~\ref{sec:pre}) that are necessary to describe our main
contribution DySECT (Section~\ref{sec:dys}).  In Section~\ref{sec:vir}
we show our in-place migration techniques.  Afterwards, we test all
hash tables on multiple benchmarks (Section~\ref{sec:exp}) and draw our
conclusion (Section~\ref{sec:con})

\section{Related Work}
\label{sec:rel}
\psfrage{I think this is too long. We do not need a philosophical reflection here. In particular when this looses the focus on space efficiency.}

The use of hash tables and other hashing based algorithms has a
long history in computer science.  The classical methods and results
are described in all major algorithm textbooks~\cite{Knuth3}.

Over the last one and a half decades, the field has regained
attention, both from theoretical and the practical point of view.  The
initial innovation that sparked this attention was the idea that
storing an element in the less filled of two ``random'' chains leads to
incredibly well balanced loads.  This concept is called the power of
two choices~\cite{ThePowerOfTwoChoicesInRandomizedLoadBalancing}.

It led to the development of cuckoo hashing~\cite{CuckooHashing}.
Cuckoo hashing extends the power of two choices by allowing to move
elements within the table to create space for new elements (see
Section~\ref{sec:pre_cuckoo} for a more elaborated explanation).
Cuckoo hashing revitalized research into space efficient hash tables.
Probabilistic bounds for the maximum fill
degree~\cite{CuckooHashingFurtherAnalysis,
  TightThresholdsForCuckooHashingViaXorSat} and expected displacement
distances~\cite{ OnTheInsertionTimeOfCuckooHashing,
  AnAnalysisOfRandomWalkCuckooHashing} are often highly non-trivial.

\psfrage{rewrote this paragraph which was misleading.}
Cuckoo hashing can be naturally generalized into two directions in order to make it more space efficient:
allowing $H$ choices \cite{SpaceEfficientWithWorstCaseConstantAccess} or extending cells in the table
to \emph{buckets} that can store $B$ elements. We will summarize this under the term \emph{bucket cuckoo hashing}.
%% Cuckoo hashing also spawned many variants that further improve its
%% performance.  The most notable one being bucket cuckoo hashing
%% ($d$-ary cuckoo hashing) a variant that groups cells together in
%% buckets thus improving cache usage, increasing displacement
%% opportunities and maximum fill
%% degree~\cite{SpaceEfficientWithWorstCaseConstantAccess,
%%   BalancedAllocationAndDictionariesWithTightlyPackedConstantSizeBin,
%%   TheRandomGraphThresholdForKOrientiabilityAndAFastAlgorithmForOptimalMultipleChoiceAllocation}
%% (see Section~\ref{sec:pre_cuckoo} for more details).

Further adaptations of cuckoo hashing include:multiple concurrent
implementations either powered by bucket locking, transactional
memory~\cite{AlgorithmicImprovementsForFastConcurrentCuckooHashing},
or fully lock-less~\cite{LockFreeCuckooHashing}; a de-amortization
technique that provides provable worst case guarantees for
insertions~\cite{DeAmortizedCuckooHashingProvableWorstCasePerformanceAndExperimentalResults,
  UsingAQueueToDeAmortizeCuckooHashingInHardware}; and a variant that
minimizes page-loads in a paged memory
scenario~\cite{CuckooHashingWithPages}.

Some non-cuckoo space efficient hash tables continue to use linear
probing variants.  \emph{Robin Hood hashing} is a technique that was
originally introduced in 1985~\cite{RobinHoodHashing}. The idea behind
Robin Hood hashing is to move already stored elements during
insertions in a way that minimizes the longest possible search
distance.  Robin Hood hashing has regained some popularity in recent
years, mainly for its interesting theoretical properties and the
possibility to reduce the inherent variance of linear probing.
\psfrage{commented out hopscotch stuff since its not space efficient and we need the space}
%% \emph{Hopscotch hashing}~\cite{HopscotchHashing} is a technique, that
%% originated in the research of concurrent hash tables, but it can also
%% manage densely filled hash tables.  Hopscotch hashing also moves
%% elements within the table to reduce the maximum search distance.
%% Additionally it introduces an acceleration data structure that is
%% designed to improve lookup performance especially in cases where the
%% searched key is not found.

All these publications show that there is a clear interest in
developing hash tables that can be more and more densely filled.
Dynamic hash tables on the other hand seem to be considered a solved
problem.  One paper that takes on the problem of dynamic hash tables
was written by Dietzfelbinger at
al.~\cite{DynamicPerfectHashingUpperAndLowerBounds}.  It predates
cuckoo hashing, and much of the attention for space efficient hashing.
All memory bounds presented are given without tight constant factors.
The lack of implementations and theory about dense dynamic hash tables
is where we pick up and offer a fast hash table implementation that
supports dynamic growing with tight space bounds.

\section{Preliminaries}
\label{sec:pre}
A hash table is a data structure for storing key-value-pairs ($\langle
key, data \rangle$) that offers the following functionality:
\verb~insert~ -- stores a given key-value pair or returns a reference\psfrage{was STL-speak ``iterator'' that only C++ programmers will understand?}
to it, if it is already contained; \verb~find~ -- given a key returns
an reference to said element if it was stored, and $\bot$ otherwise;
and \verb~erase~ -- removes a previously inserted element (if
present).

% An alternative model which is sometimes used in literature considers
% arbitrary elements instead of key value pairs.  In that model, the key
% is extracted using an extractor function.  All techniques presented in
% this paper also work in the extractor model without any adaptation.

Throughout this paper $n$ denotes the number of elements and $m$ the
number of cells ($m > n$) in a hash table.  We define the load factor
as $\delta = n/m$.  Tables can usually only operate efficiently up to a
certain maximum load factor.  Above that, operations get slower or have a
possibility to fail.  When implementing a hash table one has to decide
between storing elements directly in the hash table -- \emph{Closed
  Hashing} -- or storing pointers to elements -- \emph{Open
  Hashing}. This has an immediate impact on the amount of memory required
(\emph{closed}: $m\cdot |\textit{element}|$ and \emph{open}:
$m\cdot|\textit{pointer}|+n\cdot|\textit{element}|$).

For large elements (i.e., much larger then the size of a pointer), one
can use a non-space efficient hash table with open hashing to reduce
the relevant memory factor.  Therefore, we restrict ourselves to the
common and more interesting case of elements whose size is close
to that of a pointer.  For our experiments we use 128bit elements
(64bit keys and 64bit values).  In this case, open hashing introduces
a significant memory overhead (at least $1.5\times$). For this reason,
we only consider closed hash tables. Their memory efficiency
is directly dependent on the table's load.  To reach high fill degrees with
closed hashing tables, we have to employ \emph{open addressing}
techniques.  This means that elements are not stored in predetermined
cells, but can be stored in one of several possible places (e.g. linear
probing, or cuckoo hashing).

\subsection{\texorpdfstring{$\alpha$}--Space Efficient Hash Tables}
\myparagraph{Static.}
\label{sec:pre_staticspace}
We call a hashing technique $\alpha$-space efficient when it can work
effectively using at most $\alpha \cdot n_{\textit{curr}} \cdot
\textit{size}(\textit{element}) + O(1)$ memory. In this case we define
working efficiently as having average insertion times in
$O(\frac{1}{1-\delta})$. This is a natural estimation for insertion
times, since it is the expected number of fully random probes needed to
hit an empty cell ($1-\delta$ is the fraction of empty
cells).

%\[\alpha \cdot size_min = \alpha\cdot n\cdot size(element)\]

In many closed hashing techniques (e.g.~linear probing, cuckoo
hashing) cells are the same size as elements. Therefore, being
$\alpha$-space efficient is the same as operating with a load factor
of $\delta = \alpha^{-1}$.
%(because $\alpha\cdot n \cdot size(element) = m\cdot size(element)$).
Because of this, we will mostly talk about the load factor of a table
instead of its memory usage.\psfrage{commented out hopscotch stuff}
%% Some techniques like hopscotch hashing use additional per cell
%% information.  These hash table architectures, have to be filled more
%% densely to achieve the same space efficiency.

\myparagraph{Dynamic.}
The definition of a space efficient hashing technique given above is
specifically targeted for statically sized hash tables.  We call an
implementation \emph{dynamically $\alpha$-space efficient} if an instantiated
table can grow arbitrarily large over its original capacity while
remaining smaller than $\alpha\cdot n_{\max}\cdot
\textit{size}(\textit{element}) + O(1)$ at all times.
%\[mem_{curr} \leq \alpha\cdot max(n_{curr})\cdot size(element)\]

One problem for many implementations of space efficient hash tables is
the migration.  During a normal full table migration, both the
original table and the new table are allocated.  This requires
$m_{\textit{new}} + m_{\textit{old}}$
cells. Therefore, a normal full table migration is never more than 2-space
efficient.  The only option for performing a full table migration with
less memory is to increase the memory in-place (see
Section~\ref{sec:vir}).
Similar to static $\alpha$-space efficiency, we will mostly talk about
the \emph{minimum load factor} $\delta_{\min} = \frac{1}{\alpha}$ instead of $\alpha$.

% We note that this
% definition of dynamic $\alpha$-space efficient hash tables does not
% enforce size reduction due to the deletion of elements. However it
% enforces that cells previously used by deleted elements must be
% \emph{reused}.

\subsection{Cuckoo Hashing}
\label{sec:pre_cuckoo}
Cuckoo hashing is a technique to resolve hash conflicts in a hash
table using open addressing. Its main draw is that it guarantees
constant lookup times even in densely filled tables.  The
distinguishing technique of cuckoo hashing is that $H$ hash functions
($h_1, ... , h_H$) are used to compute $H$ independent positions. Each
element is stored in one of its positions.  Even if all positions are
occupied one can often move elements to create space for the current
element. We call this process \emph{displacing} elements.

%TODO MAYBE ALREADY GRAPH VIEW

Bucket cuckoo hashing is a variant where the cells of the hash table
are grouped into buckets of size $B$ ($m/B$ buckets).  Each element
assigned to one bucket can be stored in any of the bucket's cells.
Using buckets one can drastically increase the number of elements that
can be displaced to make room for a new one, thus decreasing the
expected length of displacement paths.

\emph{Find} and \emph{erase} operations have a guaranteed constant
running time. Independent from the table's density, there are $H$
buckets -- $H\cdot B$ cells -- that have to be searched to find an
element.

During an \emph{insert} the element is hashed to $H$ buckets.  We
store the element in the bucket with the most free space.  When all
buckets are full we have to move elements within the table such that a
free cell becomes available.

\label{sec:cuckoo_graph}
To visualize the problem of displacing elements, one can think of the
directed graph implicitly defined by the hash table.  Each bucket
corresponds to a node and each element induces an edge between
the bucket it is stored in and its $H-1$ alternate buckets.  To insert
an element into the hash table we have to find a path from one of its
associated buckets to a bucket that has free capacity.  Then we
move elements along this path to make room in the initial bucket.
The two common techniques to find such paths are \emph{random walks} and
\emph{breadth first searches}.

\section{DySECT (Dynamic Space Efficient Cuckoo Table)}
\label{sec:dys}
A commonly used growing technique is to double the size of a hash
table by migrating all its elements into a table with twice its
capacity.  This is of course not memory efficient.  The idea behind
our dynamic hashing scheme is to double only parts of the overall
data structure.  This increases the space in part of our
data structure without changing the rest.  We then use cuckoo
displacement techniques to make this additional memory reachable from
other parts of the hash table.

\subsection{Overview}
\begin{figure*}[ht]
  \centering
  \includegraphics[height=5.3cm]{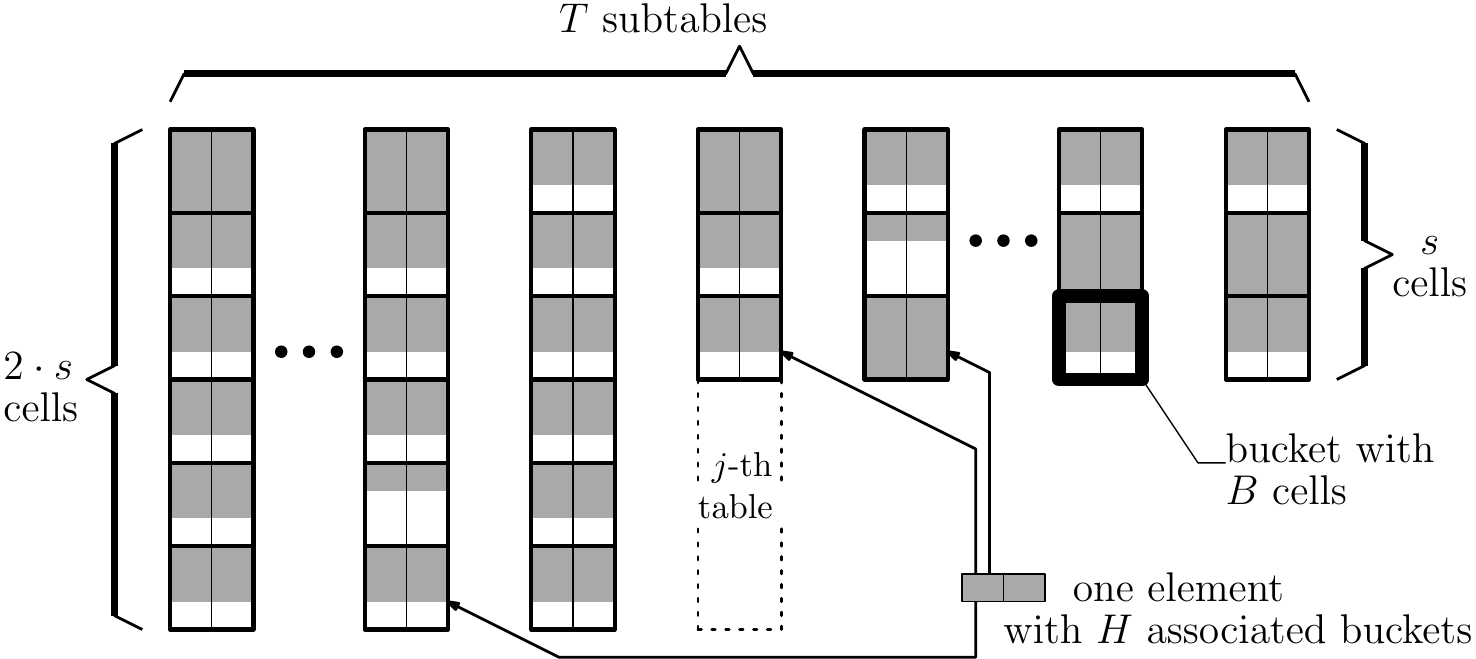}
  \caption{\label{fig:structure} Schematic Representation of a DySECT Table. }
\end{figure*}

Our DySECT hash table consists of $T$ subtables (shown in
Figure~\ref{fig:structure}) that in turn consist of buckets, which can
store $B$ elements each.  Each element has $H$ associated buckets --
similar to cuckoo hashing -- which can be in the same or in different
subtables.  $T$, $B$, and $H$ are constant that will not change during
the lifetime of the table.  Additionally, each table is initialized
with a minimum fill ratio $\delta_{\min}$. The table will never exceed
$\delta_{\min}^{-1}\cdot n$ cells once it begins to grow over its
initial size.

To find a bucket associated with an element $e$, we
compute $e$'s hash value using the appropriate hash function
$h_i(e)$. The hash is then used to compute the subtable and the bucket
within that subtable.
To make this efficient we use powers of two for the number of
subtables ($T = 2^t$), as well as for the number of buckets per subtable (subtable size $s =
2^x\cdot B$).  Since the number of subtables is constant, we can use the
first $t$ bits from the hashed key to find the appropriate subtable.
From the remaining bits we compute the bucket within that subtable using
a bitmask ($h_i(e)\,\texttt{\&}\,(2^x-1) = h_i(e) \mod 2^x$).

%% We split that hash value into two parts, then we use the
%% first part to chose a subtable, and the second part to
%% choose one of the contained buckets.

%% To understand the performance of this hash table, it is important to
%% realize that when the hash table is used regularly, the first level
%% table will always remain cached. This is important, because otherwise lookups
%% would cause unnecessary cache misses that would not have happened in a
%% single level hash table.  Because of this we expect the average number
%% of cache misses during hash table operations to be similar to those of
%% a normal cuckoo hash table.

\subsection{Growing}
As soon as the (overall) table contains enough elements such that the
memory constraint can be kept during a subtable migration, we grow one
subtable by migrating it into a table twice its size.  We migrate
subtables in order from first to last. This ensures that no subtable
can be more than twice as large as any other.

Assume that we have $j$ large subtables ($2 s$) than $m = (T+j)\cdot
s$. When $\delta_{\min}^{-1}\cdot n > m+2s$ we can grow the first
subtable while obeying the size constraint (the newly allocated table
will have $2 s$ cells).  Doubling the size of a subtable increases the
global number of cells from $m_{\textit{old}} = (T+j)\cdot s$ to
$m_{new} = m_{old}+s = (T+j+1)\cdot s$ (grow factor
$\frac{T+j+1}{T+j}$).  Note that all subsequent growing operations
migrate one of the smaller tables until all tables have the same
size.  Therefore, each grow until then increases the overall capacity
by the same absolute amount (smaller relative to the current size).

The cost of growing a subtable is amortized by all insertions since
the last subtable migration.  There are $\delta_{\min} \cdot s =
\Omega(s)$ insertions between two migrations.  One migration takes
$\Theta(s)$ time.  Apart from being amortized, the migration is cache
efficient since it accesses cells in a linear fashion.  Even in the
target table cells are accessed linearly.  We assign elements to
buckets by using bits from their hash value. In the grown table we use
exactly one more bit than before (double the number of buckets).  This
ensures that all elements from one original bucket are split between
two buckets in the target table.  Therefore no bucket can overflow
and no displacements are necessary.

In the implicit graph model of the cuckoo table
(Section~\ref{sec:cuckoo_graph}), growing a subtable is equivalent to
splitting each node that represents a bucket within that subtable. The
resulting graph becomes more sparse, since the edges (elements) are
not doubled, making it easier to insert subsequent elements.

\subsection{Shrinking}
% In many use cases automatic shrinking is not necessary.  It worsens
% performance by taking time for the migration and making the remaining
% table more dense.  Therefore, we forgo automatic shrink operations
% whenever enough elements are removed from the table.
If shrinking is necessary it can work similarly to growing. We
replace a subtable with a smaller one by migrating elements from one
to the other.  During this migration we join elements from two buckets
into one. Therefore it is possible for a bucket to overfill.  We
reinsert these elements at the end of the migration.  Obviously, this
can only affect at most half the migrated elements.

When automatically triggering the size reduction, one has to make sure
that the migration cost is amortized. Therefore, a grow operation
cannot immediately follow a shrink operation.  When shrinking is
enabled we propose to shrink one subtable when $\delta_{\min}^{-1}\cdot n <
m-s'$ elements ($s'$ size of a large table, $m_{\textit{new}} =
m_{\textit{old}} -s'/2$).  Alternatively, one could implement a
\emph{shrink to size} operation that is explicitly called by the user.

% Notice that the allocated memory actually increases during the
% migration itself.

\subsection{Difficulties for the Analysis of DySECT}
There are two factors specific to DySECT impacting its performance:
\emph{inhomogeneous table resolution} and \emph{element
  imbalance}.

\myparagraph{Imbalance through Inhomogeneous Table Resolution.}
\label{sec:inhom_res}
By growing subtables individually we introduce a size imbalance
between subtables. Large subtables contain more buckets but the number
of elements hashed to a large subtable is not generally higher than
the number of elements that are hashed to a small subtable.  This
makes it difficult to spread elements evenly among buckets.
Imbalanced bucket fill ratios can lead to longer insertion times.

Assume there are $n$ elements in a hash table with $T$ subtables, $j$
of which have size $2s$ the others have size $s$. If elements are
spread equally among buckets then all small tables have around
$n/(T+j)$ elements, and the bigger tables have $2n/(T+j)$ elements.
For each table there are about $Hn/T$ elements that have an associated
bucket within that table.  This shows that having more hash functions
can lead to a better balance.

For two hash functions ($H=2$) and only one grown table ($j=1$) this
means that $\approx 2n/(T+1)$ elements should be stored in the first
table to achieve a balanced bucket distribution.  Therefore, nearly
all elements associated with a bucket in the first table ($\approx
2n/T$) have to be stored there.  This is one reason why $H=2$ does not
work well in practice.

\myparagraph{Imbalance through Size Changes.}
In addition to the problem of inhomogeneous tables there is an
inherent balancing problem introduced by resizing subtables. It is
clear that a newly grown table is not filled as densely as other
tables.  Since we double the table size, grown tables can only be
filled to about 50\%.

Assume the global table is filled close to 100\% when the first table
grows.  Now there is capacity for $s$ new elements but this
capacity is only in the first table, elements that are not hashed to
the first table, automatically trigger displacements leading to slow
insertions.  Notice that repeated \verb~insert~ and \verb~erase~ operations help to
equalize this imbalance, because elements are more likely inserted
into the sparser areas, and more likely to be deleted from denser
areas.

% \subparagraph*{Population Density}
% Measuring and comparing the performance of space efficient growing
% tables to their statically sized counterparts is difficult.  Inserting
% elements into a densely filled table takes longer, than into an empty
% one.  Therefore, a table that is always densely filled will naturally
% be slower than one that only fills up towards the end of the
% execution -- even when ignoring eventual growing costs.

\subsection{Implementation Details}
\label{sec:dys_imp}
For our experiments (Section~\ref{sec:exp}) we use three hash
functions ($H=3$) and a bucket size of ($B=8$). These values have
consistently outperformed other options both in maximum load factor
and in \verb~insert~ performance (see Appendix~\ref{app:param}).  $T$ is
set to 256 subtables for all our tests.  To find displacement
opportunities we use breadth first search.  In our tests it performed
better than random walks, since it better uses the read cache lines
from one bucket.

The hash table itself is implemented as a constant sized array of
pointers to subtables. We have to lookup the corresponding pointer
whenever a subtable is accessed.  This does not impact performance
much since all subtable pointers will be cached -- at least if the
hash table is a performance bottleneck.

\myparagraph{Reducing the Number of Computed Hash Functions.}
Evaluating hash functions is expensive, therefore, reducing the number
of hash functions computed per operation can increase the performance
of the table.  The hash function we use computes 64bit hash values
(i.e. xxHash\footnote{\url{xxhash.com}}).  We split the 64bit hash
value into two 32bit values. All common bucket hash table sizes can be
addressed using 32 bits (up to $2^{32}$ buckets $2^{35} \approx 34$
billion elements consuming 512GiB memory).

When $H > 2$ we can use \emph{double
  hashing}~\cite{TheAnalysisOfDoubleHashing,
  LessHashingSamePerformance} to further reduce the number of computed
hash functions. Double hashing creates an arbitrary number of hash
values using only two original hash functions $h'$ and $h''$.  The
additional values are linear combinations computed from the original
two values, $h_i(key) = h'(key) + i\cdot h''(key)$.

Combining both of these techniques, we can reduce the number of
computed hash functions to one 64bit hash function.  This is
especially important during large displacements where each
encountered element has to be rehashed to find its alternative
buckets.

\section{(Ab)Using Virtual Memory}
\label{sec:vir}
In this section we show how one can use virtual memory and memory
overcommitting, to eliminate the indirections from a DySECT hash table.
The same technique also allows us to implement hash tables that can
grow using an in-place full table migration.  If we grow these tables
in small increments, they can grow while enforcing a strict size
constraint.

To explain these techniques, we first have to explain how to use memory
overcommitting and virtual memory to create a piece of memory that can
grow in-place.
Note that this technique violates best programming
practices and is not fully portable to some systems.

The idea is the following: the operating system will -- if configured
to do so -- allow memory allocations larger than the machine's main
memory, with the anticipation that not all allocated memory will
actually be used.  Only memory pages that are actually used will be
mapped from virtual to physical memory pages.  Thus, for the purpose
of space efficiency the memory is not yet used.  Initializing parts of
this memory is similar to allocating and initializing new memory.

\subsection{Improving DySECT}
\label{sec:vir_dysect}
Accessing a DySECT subtable usually takes one indirection.  The
pointer to the subtable has to be read from an array of pointers
before accessing the actual subtable.  Instead of using an array of
pointers, we can implement the subtables as sections within one large
allocation (size $u$).  We choose $u$ larger than the actual main
memory, to allow all possible table sizes.  This has the advantage
that the offset for each table can be computed quickly ($t_i =
\frac{u}{T}\cdot i$), without looking it up from a table.

The added advantage is that we can grow subtables in-place.  To
increase the size of a subtable, it is enough to initialize a
consecutive section of the table (following the original subtable).
Once this is done, we have to redistribute the table's elements.  This
allows us to grow a subtable without the space overhead of
reallocation.  Therefore, we can grow earlier, staying closer to the
minimum load factor $\delta_{\min}$.  The in-place growing mechanism is easy in this
case, since the subtable size is doubled.

% Approximately half the elements are already in the correct bucket,
% since the number of buckets is a power of two and we use a bitmask to
% distribute elements (the grown table uses one more bit).  All other
% elements have to be moved into one of the newly initialized buckets.

\subsection{Implementing other size constrained tables}
\label{sec:vir_competitors}
Similarly to the technique above, we can implement any hash table
using a large allocation, initializing only as much memory as the
table initially needs.  The used hash table size can be increased
in-place by initializing more memory.  To use this additional memory
for the hash table, we have to perform an in-place migration.

To implement fast in-place migration, we need the correct addressing
technique.  There are two natural ways to map a hash value $h(e)$ to
a cell in the table (size $s$).  Most programmers would use a slow
modulo operation ($h(e)\mod s$).  This is the same as using the least
significant digits when addressing a table whose size is a power of
two.  A better way is to use a scale factor ($\lfloor h(e)\cdot
\frac{s}{max(h)}\rfloor$). This is similar to using the most
significant bits to address a table whose size is a power of two.  The
second method has two important advantages, it is faster to compute
and it helps to make the migration cache efficient.
When we use the second method the elements in the hash table are close
to being sorted by their hash value (in the absence of collisions
they would be sorted).

The main idea of all our in-place migration techniques is the
following. If we use a scale factor for our mapping -- in the new table -- most elements will
be mapped to a position that is larger than their position in the old
table.  Therefore, rehashing elements starting from the back of the
original table creates very few conflicts.  Elements that are mapped
to a position earlier than their current position are buffered and
reinserted at the end of the migration.  When using this technique,
both the old and the new table, are accessed linearly in reverse
order (from back to front).  Making the migration cache efficient and
easy to implement.

For a table that was initialized with a min load factor
$\delta_{\min}$ we trigger growing once the table is loaded more than
$\frac{\delta_{\min} + 1}{2}$. We then increase the capacity $m$ to
$\delta_{\min}^{-1}\cdot n$.  Repeated migrations with small growing
amounts are still inefficient, since each element has to be moved.
%Therefore, one cannot hope that these variants outperform a DySECT
%table when repeated growing in small increments is necessary.

This blueprint can be used, to implement in-place growing variants of
most if not all common hashing techniques.  We used these same ideas
to implement variants of \emph{linear probing}, \emph{robin hood
  hashing}, and \emph{bucket cuckoo hashing}.  Although some variants
have their own optimized migration.  Robin Hood hashing can be adapted
such that the table is truly sorted by hash value (without much
overhead) making the migration faster than repeated reinsertions.
Bucket cuckoo hashing has a somewhat more complicated migration
technique, since each element has multiple possible positions, and one
bucket can overflow.  The best strategy here is to try to reinsert
each element with the hash function ($h_1,...,h_H$) that was
previously used to store it.

\section{Experiments}
\label{sec:exp}
There are many factors that impact hash table performance.  To show
that our ideas work in practice we use both micro-benchmarks and
practical experiments.

All reported numbers are averaged by running each experiment five
times.  The experiments were executed on
a server with two Intel Xeon E5-2670 CPUs (2.3GHz base frequency)
and 128GB RAM (using gcc 6.2.0 and Ubuntu 14.04).%
\footnote{Experiments on a desktop machine yielded similar results.}
%% two machines: a two socket
%% server computer with Intel Xeon E5-2670 CPUs (2.3GHz base frequency)
%% and 128GB RAM (using gcc 6.2.0 and Ubuntu 14.04); and a Core\psfrage{discuss: needed?} i7-4790T
%% (2.7GHz base frequency) desktop machine with 16GB RAM (using gcc 6.3.0
%% and Debian unstable).  The results were qualitatively the same,
%% therefore, we only report numbers from the server computer.

To put the performance of our DySECT table into perspective, we
implement and test several other options for space efficient hashing
using the method described in Section~\ref{sec:vir_competitors}.  We
use our own implementations, since no hash table found online supports
our strict space-efficiency constraint.
With the technique described in Section~\ref{sec:vir_competitors}, we
implement and test hash tables with \emph{linear probing}, \emph{robin
  hood hashing}, and \emph{bucket cuckoo hashing} (similar to DySECT
we choose $B=8$ and $H=3$ see Appendix~\ref{app:param} for experiments with other parameter settings).  For each table, we implemented an individually tuned
cache efficient in-place migration algorithm.\psfrage{describe in appendix?}
\psfrage{commented out hopscotch stuff}
%% We also implemented a similar variant of \emph{hopscotch
%%   hashing}. During our tests, we found that we need neighborhoods with
%% 64 or more cells to even reach filling ratios of 90\% (of all cells).
%% Each cell in a hopscotch table stores a $|$neighborhood$|$ bits of
%% acceleration data in addition to its contents.  This bounds the
%% effective filling degrees to $\leq\frac{2}{3}$. Making hopscotch
%% hashing unfeasible for the purpose of $\alpha$-space efficient
%% hashing. Therefore, we omit hopscotch hashing from our experimental
%% evaluation

\myparagraph{Without Virtual Memory/ Memory Overcommitting.}  All
implementations described above work with the trick described in
Section~\ref{sec:vir}.  The usefulness of this technique is arguable,
since abusing the concept of virtual memory in this way is problematic
not only from a software design perspective.  It directly violates
best practices, and reduces portability to many systems.  The only
table that can achieve dynamic $\alpha$-space efficiency without this
technique is our DySECT hash table.  It is notable that this
implementation is never significantly worse than DySECT with
overcommitting (this variant is displayed using a dashed line).

For each competitor table, we also implemented a variant that uses
subtables combined with normal migrations (small grow factor, similar
to in-place variants).  Elements are first hashed to subtables and
then hashed within that table.  They cannot move between
subtables. These variants are not strictly space efficient.  The
subtables are generally small ($n/T$), therefore migrations will
usually not violate the size constraint (too much).  There can be
larger subtables, since imbalances between subtables cannot be
regulated.  Throughout this section, we display these variants with
dashed lines (similar to the DySECT variant without memory
overcommitting).

\subsection{Influence of Fill Ratio (Static Table Size)}
\label{sec:exp_eps}

\begin{figure*}[ht]
  \centering
  \includegraphics[width=\textwidth]{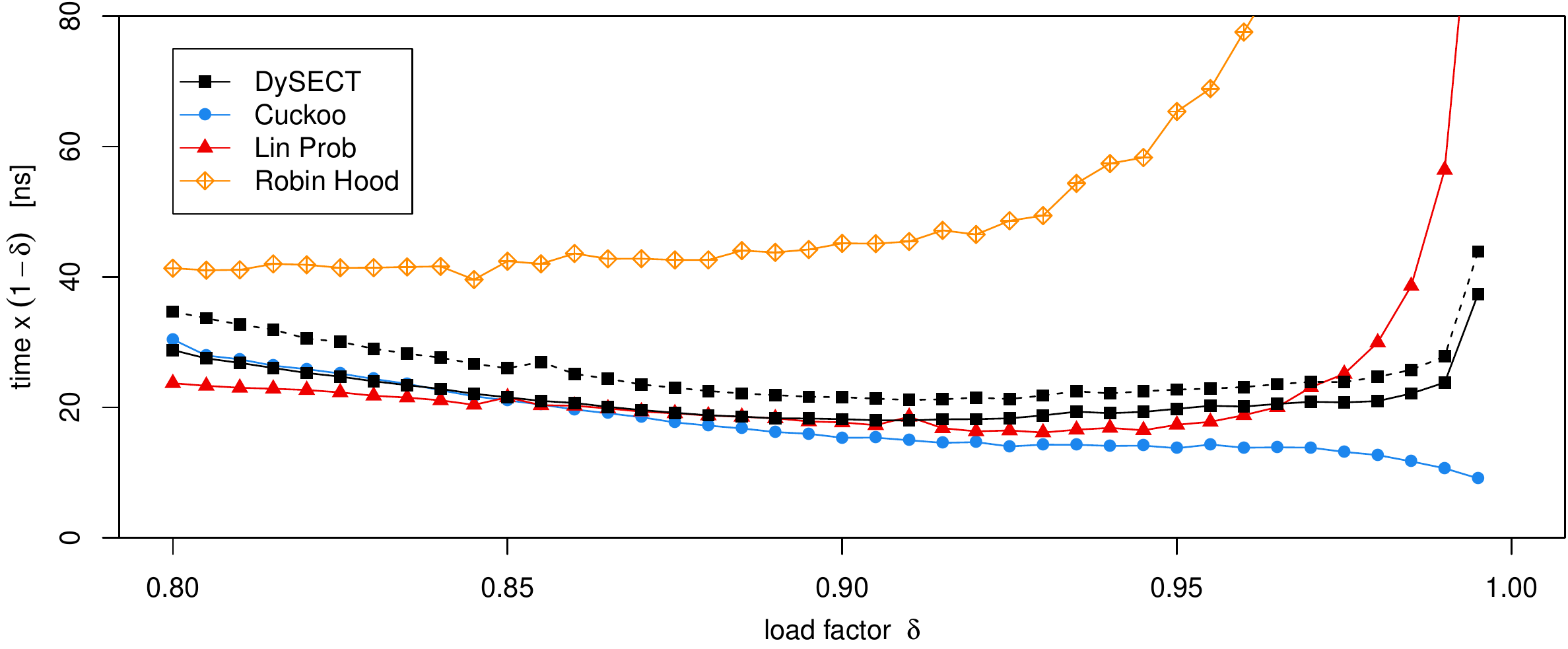}
  \caption{\label{fig:eps_insert} Insertions into a Static Table.
    Here we show the influence from the load factor, on the
    performance of insertions.  To make insertion time more readable, we normalize it with
    $t_{op}\cdot (1-\delta)$. }
\end{figure*}

\begin{figure}[ht]
  \centering
  \includegraphics[width=\columnwidth]{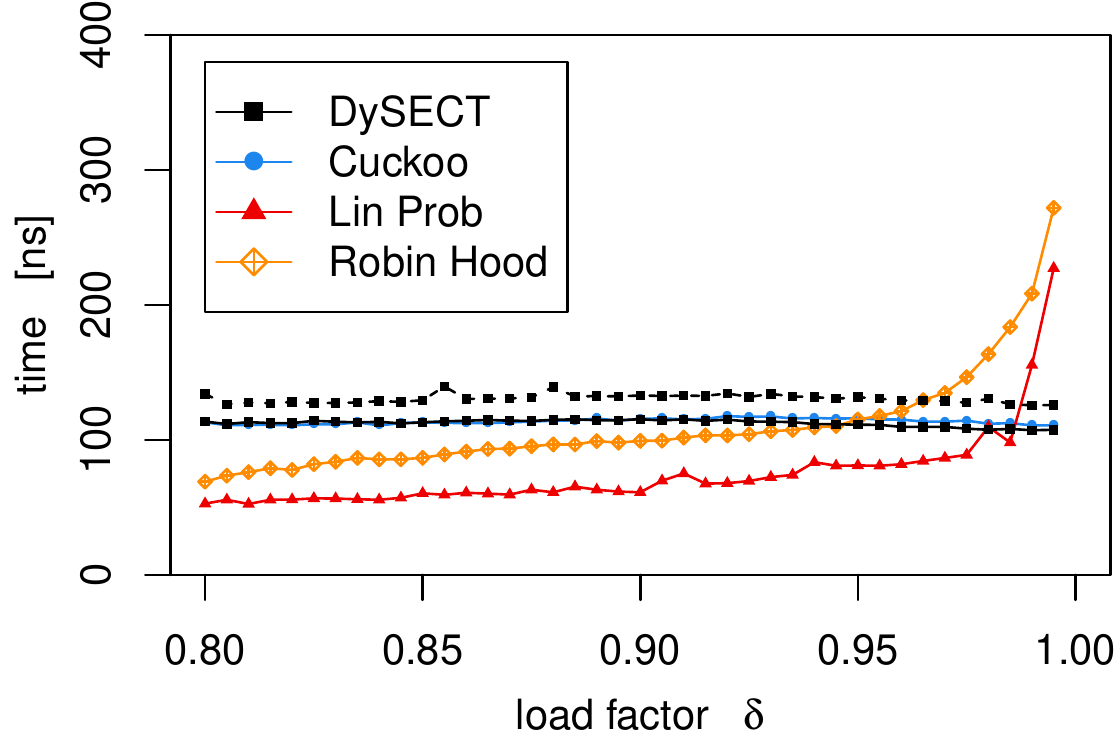}
  \caption{Performance of Successful Finds.  DySECT's find performance is independent from the load factor. }
  \label{fig:eps_finds}
\end{figure}

\begin{figure}[ht]
  \centering
  \includegraphics[width=\columnwidth]{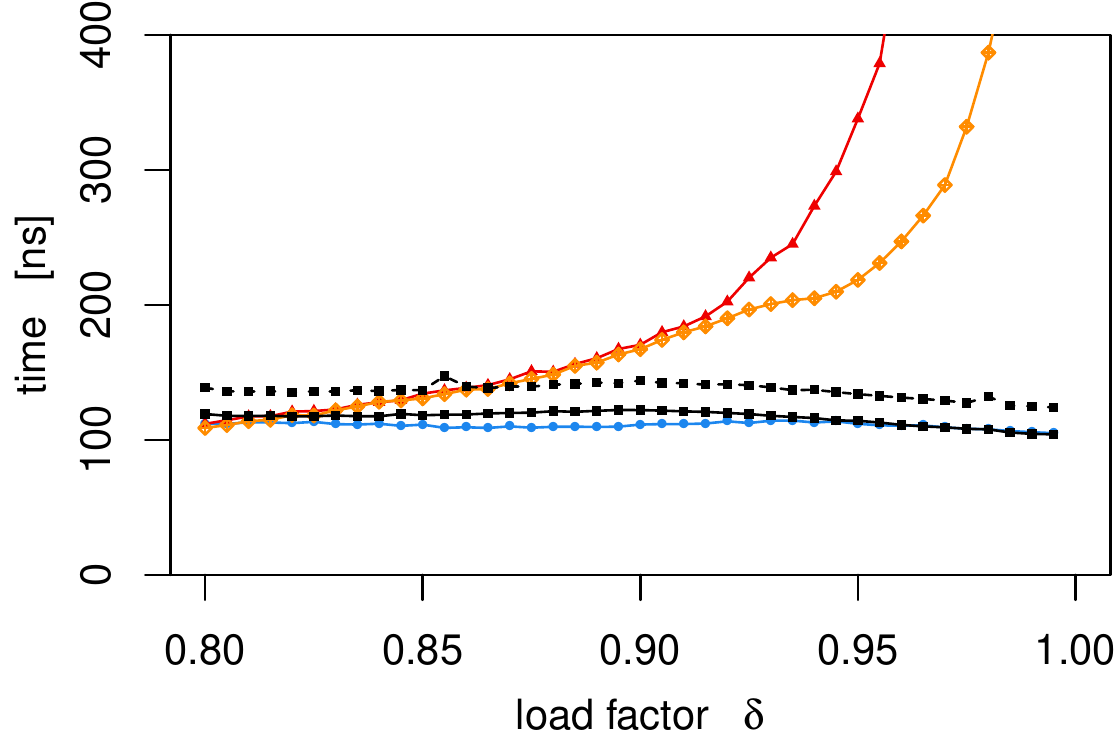}
  \caption{Performance of Unsuccessful Finds.  DySECT's find performance is independent from the operations success. }
  \label{fig:eps_findu}
\end{figure}

The following test was performed by initializing a table with
$m\approx25\,000\,000$~cells (non-growing). Then elements are inserted
until there is a failing insertion.  At
different stages, we measure the running time of new insertion
(Figure~\ref{fig:eps_insert}), and find (Figure~\ref{fig:eps_finds} and \ref{fig:eps_findu})
operations (averaged over 1000 operations).  Finds are measured using
either randomly selected elements from within the table (successful),
or by searching random elements from the whole key space
(unsuccessful).  We omit testing multi table variants of the
competitor tables. They are not suitable for this test since forcing a
static size limits the possibility to react to size imbalances between
subtables (in the absence of displacements).

As to be expected, the insertion performance of depends
highly on the fill degree of the table.  Therefore, we show it
normalized with $\frac{1}{1-\delta}$ which is the expected number
of fully random probes to find a free cell and thus a natural estimate
for the running time. We see that -- up to a certain point -- the
insertion time behaves proportional to $\frac{1}{1-\delta}$ for all
tables. Close to the capacity limit of the table, the insertion time increases sharply.
DySect has a smaller capacity limit than cuckoo due inhomogeneous table resolution (see
Section~\ref{sec:inhom_res}).\psfrage{commented out lengthy discussion of $1/T$ effects which rather illuminate a flaw in our experiment than an important insight.}
%% When using our DySECT table, the running time remains stable for longer than when we use linear probing but
%% not as long, as for classic cuckoo hashing.  One reason for this is
%% that the DySECT table consists of $T$ subtables with power of two sizes
%% and thus does not allow exactly 25\,000\,000 cells.  It contains
%% 30\,784 cells less than the other tested tables.  Therefore, our table is
%% slightly more full.  Another reason is that in order to get close to
%% 25\,000\,000 cells, we have 125 large subtables (twice the size of the
%% others).  This introduces inhomogeneous table resolution (see
%% Section~\ref{sec:inhom_res}), which decreases our performance compared
%% to cuckoo hashing.

Figure~\ref{fig:eps_finds} and \ref{fig:eps_findu} show the performance of find operations.
Linear probing performs relatively well on successful find operations,
up to a fill degree of over 95\%.  The reason for this is that many
elements were inserted into the table when the table was still
relatively empty.  They have very short search distances, thus
improving find performance.  Successful find performance can still be
an issue in applications.  An element that is inserted when the table
is already decently filled can have an extremely long search distance.
This leads to a high running time variance on find operations.
Unsuccessful finds perform really badly, since all cells until the
next free cell have to be probed.  Their performance is much more
related to the filling degree of the hash table.
Robin Hood hashing performs somewhat similar to linear probing.  It
worsens the successful find performance by moving previously inserted
elements from their original position, in order to achieve better
unsuccessful find performance on highly filled tables.  Overall, Robin
Hood hashing is objectively worse than both DySECT and classic cuckoo hashing.
Cuckoo hashing and its variants like DySECT have
guaranteed constant running times for all find operations --
independent of their success and the table's filling degree.

\subsection{Influence of Fill Ratio (Dynamic Table Size)}
\label{sec:exp_ti}

\begin{figure*}[ht]
  \centering
  \includegraphics[width=\textwidth]{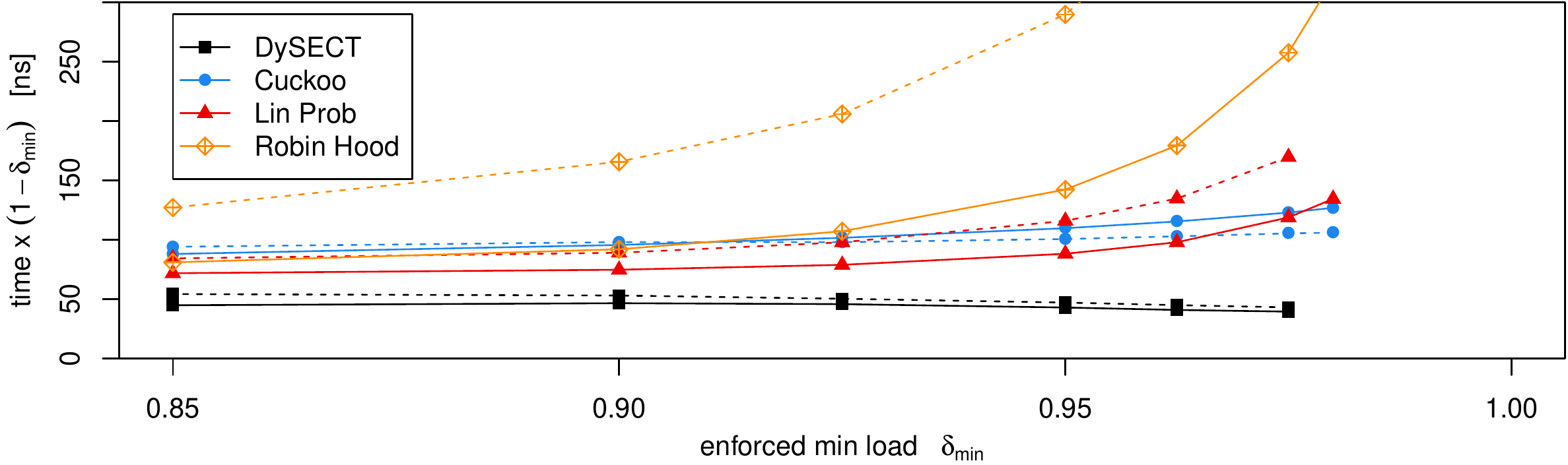}
  \caption{\label{fig:ti_insert} Insertions into a dynamic growing table
    enforcing a minimum load factor $\delta_{\min}$. }
\end{figure*}

In this test 20\,000\,000 elements are inserted into an initially
empty table.  The table is initialized expecting 50\,000 elements,
thus growing is necessary to fit all elements.  The tables are
configured to guarantee a load factor of at least $\delta_{\min}$ at all
times. Figure~\ref{fig:ti_insert} shows the performance in relation to
the load factor.  Insertion times are computed as average of all
20\,000\,000 insertions. They are normalized similar to
Figure~\ref{fig:eps_insert} (divided by $\frac{1}{1-\delta_{\min}}$).

We see that DySECT performs by far the best even with less filled
tables at 85\% load.  Here we achieve a speedup of
1.6\psfrage{switched to speedup factors which look better imho.} over
the next best solution (299ns vs. linear probing 479ns).  On denser
instances with 97.5\% load, we can increases these speedup to 2.7
(1580ns vs Cuckoo with subtables 4210ns).\psfrage{speedup among
  portable methods even higher? give it?}  With growing load, we see
the insertion times of our competitors degrade.  The combination of
long insertion times, and frequent growing phases slows them down.
There are only few insertions between two growing phases.  Making the
amortization of each growing phase challenging since each growing
phase has to move all elements. Any growing technique that uses a less
cache efficient migration algorithm, would likely perform
significantly worse.  DySECT however remains close to
$O(\frac{1}{1-\delta_{\min}})$ even for fill degrees up to 97.5$\%$.
This is possible, because only very few elements are actually touched
during each subtable migration ($\approx \frac{n}{T}$).

We also measured the performance of find operations on the created
tables, they are similar to the performance on the static table in
Section~\ref{sec:exp_eps} (see Figure~\ref{fig:eps_finds} and \ref{fig:eps_findu}), therefore,
we omit displaying them for space reasons.

\subsection{Word Count -- a Practical use Case}
\label{sec:exp_wordcount}

\begin{figure*}[ht]
  \centering
  \includegraphics[width=\textwidth]{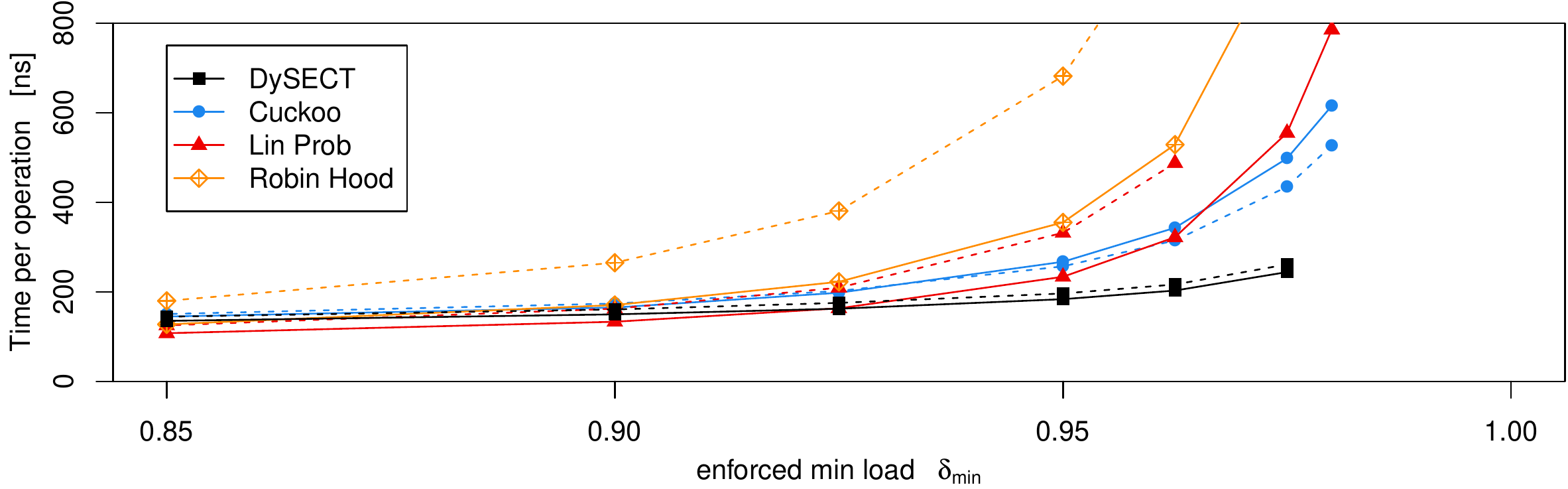}
  \caption{\label{fig:crawl} Word Count Benchmark.  The benchmark
    behaves like a mix of insert and find operations.  DySECT's
    performance is nearly independent form the load factor. }
\end{figure*}

Word count and other aggregation algorithms are some of the most
common use cases for hash tables.  Data is aggregated according to its
key, which in our case is a hash of the contained word. This is a
common application, in which static hash tables can never be space
efficient, since the final size of the hash table is usually unknown.
Here we use the first block of the \emph{CommonCrawl} dataset
(\url{commoncrawl.org/the-data/get-started}) and compute a word count
of the contained words, using our hash tables.  The chosen block has
4.2GB and contains around 240\,000\,000 words, with around
20\,000\,000 unique words. For the test, we hash each word to a 64 bit
key and insert it together with a counter.  Subsequent accesses to the
same word increase this counter.  Similar to the growing benchmark, we
start with an empty table initialized for 50\,000 elements.
%% Looking at these numbers, it becomes obvious why estimating
%% the number of unique words might be difficult.

The performance results can be seen in Figure~\ref{fig:crawl}\psfrage{make figure more narrow and put caption to the side of the figure?}.  We do
not use any normalization since each word is repeated 12 times (on
average).  This means that most operations will actually behave more
like successful find operations instead of inserting an element.  When
using our DySECT table, the running time seems to be nearly
independent from the fill degree.  We experience little to no slowdown until around 97\%.  The tables using full table migration however
become very inefficient on high load degrees\psfrage{report some speedup factors?}.

For high load factors, the performance closely resembles that of the
insertion benchmark (Figure~\ref{fig:ti_insert}).  This indicates that
\verb~insert~ performance can dominate running times even in
\verb~find~ intensive workloads.  To confirm this insight, we
conducted some experiments with mixed operations (\verb~insert~ and
\verb~find~; \verb~insert~ and \verb~erase~).  They showed, that on a
hash table with $\delta_{\min}=0.95$ DySECT outperforms linear probing
for any workload containing more than 5$\%$ insertions (see
Section~\ref{sec:exp_mix}).

\subsection{Mixed Workloads}
\label{sec:exp_mix}

\begin{figure}[ht]
  \centering
  \includegraphics[width=\columnwidth]{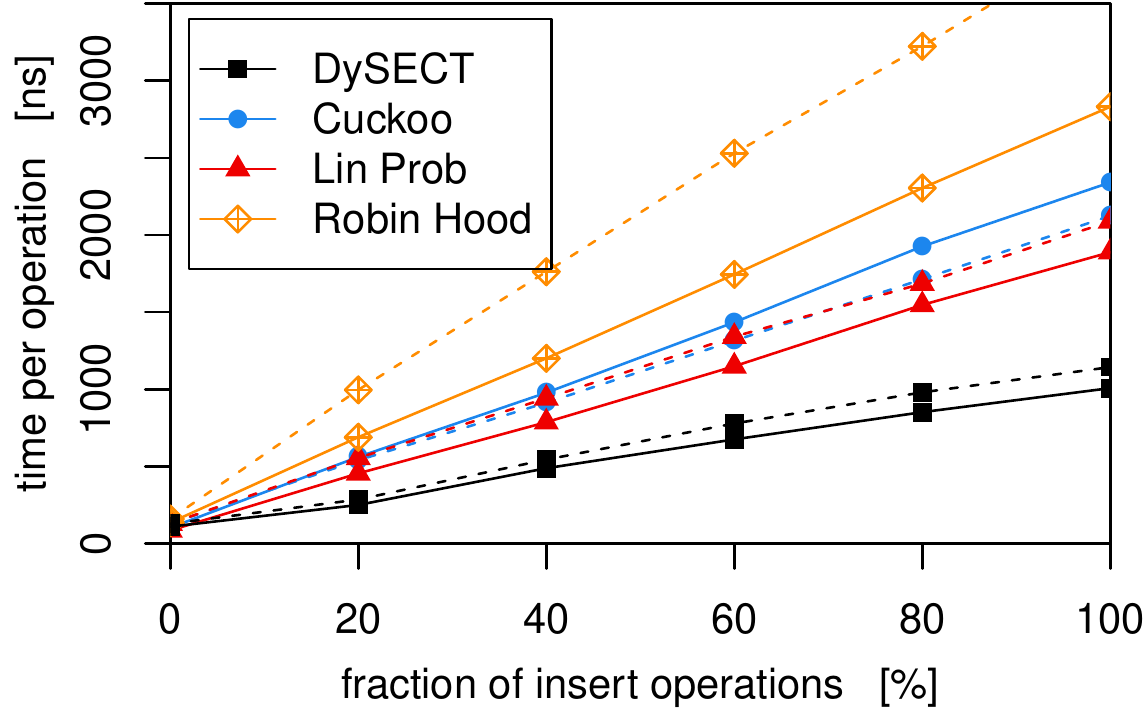}
  \caption{\label{fig:mix} Workload combining insertions with
    \texttt{finds}.}
\end{figure}

\begin{figure}[ht]
  \centering
  \includegraphics[width=\columnwidth]{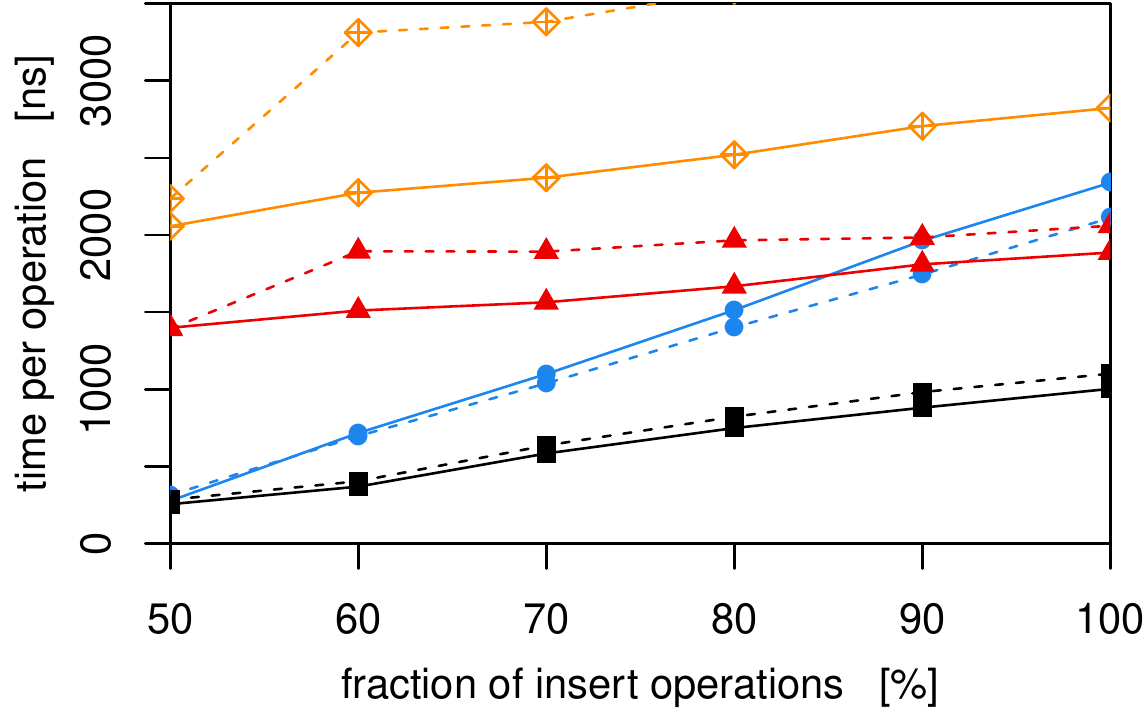}
  \caption{\label{fig:mixd} Workloads combining insertions with \texttt{erase}.}
\end{figure}

In this test, we show how the hash tables behave under mixed
workloads.  For this test we fixed the minimum load factor to
$\delta = 0.95$.  The test starts with a filled table
containing 15\,000\,000 elements (95\% filled).  On this table we perform
10\,000\,000 operations mixed between \verb~insert~ and
\verb~find~/\verb~erase~ operations.

As one might expect, the running time of a mixed work load can be
estimated with a linear combination of the used operations.

In the \verb~insert/find~ benchmark (Figure~\ref{fig:mix}), DySECT outperforms all other hash
tables.  This shows that fast insertions are important even in find
intensive workloads.  This is even more accentuated by the fact that
all performed finds are successful finds which have usually
better performance on hash tables using linear probing
The measurements with deletions (Figure~\ref{fig:mixd}) show that deletions in linear probing
tables are significantly slower than those in cuckoo tables. This
makes sense, since linear probing tables have to move elements, to fix
their invariants while cuckoo tables have guaranteed constant
deletions.

\subsection{Investigating Maximum Load Bounds}
\begin{figure}[ht]
  \centering
  \includegraphics[width=\columnwidth]{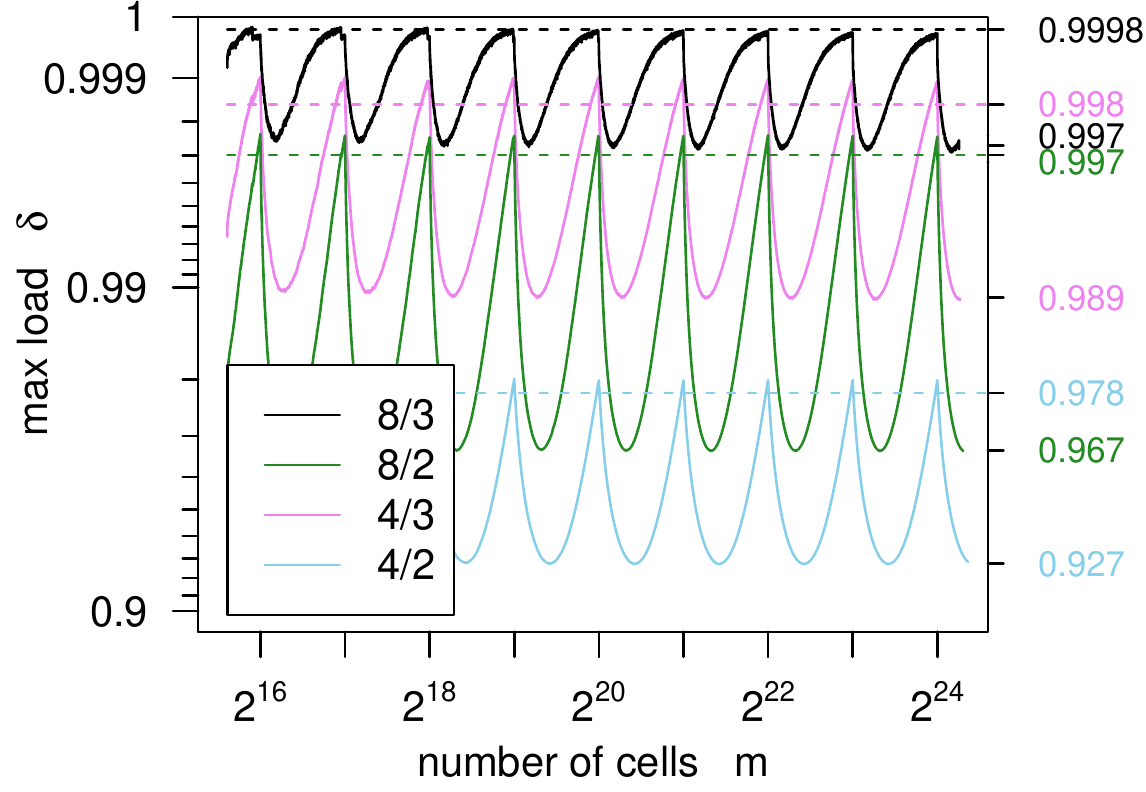}
  \caption{\label{fig:mx_load} Maximum load bounds for different
    parameterizations ($B/H$) over a varying table size. Cuckoo's bound is
    indicated by a dashed line.}
\end{figure}

We designed the following experiment, to give an indication for the
theoretical load bounds DySECT is able to support.  To do this, we
configured DySECT to grow only if there is an unsuccessful insertion
-- in our other tests the table grows in anticipation once the
size-constraint allows it.  To get even closer to the theoretical
limits, we use $T=4096$ subtables such that each grow has a smaller
relative grow factor. Additionally we increase the number of probes
used to find a successful displacement path to 16384 probes
(from 1024).

We test different combinations of bucket size ($B$) and number of hash
functions ($H$).  The test consists of inserting 20\,000\,000 elements
into a previously empty table initialized with 50\,000 cells.
Whenever the table has to grow we note its effective load (during the
migration). The maximum load bound depends on the number of large
subtables. Therefore, we show the achieved load factors over the table
capacity ($m$).  To give a perspective, of static table performance we
show dashed lines with the performance of a cuckoo hash table.  We
measured cuckoo's performance by initializing a table with
20\,000\,000 cells and filling it until there was an error (using the
same search distance of 16384).

Figure~\ref{fig:mx_load} clearly shows the cyclical nature of DySECT's
maximum load bound.  The table can be filled the most, when its
capacity is close to a power of two.  Then, all subtables have the
same size and the table behaves similar to a static cuckoo table of
the.  On some sizes DySECT even seems to outperform cuckoo hashing.
We are not sure why this is the case.  We suspect that the bound
measured for cuckoo hashing is not tight (cuckoo hashing bounds are
only measured on one table size).  This effect does not appear for the
$8/3$ parameterization that we use throughout the paper.

Both parameterizations that are using only two hash functions ($8/2$
and $4/2$) have a higher dependency on the table size.  The reason for
this is that additional hash functions help to reduce the imbalance
introduced by varied subtable resolution described in
Section~\ref{sec:inhom_res}.  Overall, we reach load bounds that are
close to those of static cuckoo hashing.

\section{Conclusion}
\label{sec:con}
We have shown that dynamically growing hash tables can be implemented
to always consume space close to the lower bound. We find it
surprising that even our simple solutions based on linear probing seem
to be new. DySECT is a sophisticated solution that exploits the
flexibility offered by bucket cuckoo hashing to significantly decrease
the number of object migrations over more straightforward approaches.
When very high space efficiency is desired, it is up to 2.7 times
better than simple solutions.

For future work, a theoretical analysis of DySECT looks
interesting. We expect that techniques previously used to analyze
bucket cuckoo hashing will be applicable in principle. However, the
already very complex calculations have to be generalized to take all
possible ratios of small versus large subtables into account.
Even for the static case and classical bucket cuckoo hashing, it is a
fascinating open question whether the observed proportionality of
insertion time to $1/(1-\delta)$ can be proven. Previous results on
insertion time show much more conservative bounds \cite{SpaceEfficientWithWorstCaseConstantAccess,OnTheInsertionTimeOfCuckooHashing,AnAnalysisOfRandomWalkCuckooHashing,BalancedAllocationAndDictionariesWithTightlyPackedConstantSizeBin}.

On the practical side, DySECT looks interesting for concurrent hashing
\cite{AlgorithmicImprovementsForFastConcurrentCuckooHashing,LockFreeCuckooHashing}
since it grows only small parts of the table at a time.
%% or a table with de-amortized
%% insertions~\cite{UsingAQueueToDeAmortizeCuckooHashingInHardware,
%%   DeAmortizedCuckooHashingProvableWorstCasePerformanceAndExperimentalResults}.

%%
%% Bibliography
%%

%% Either use bibtex (recommended),

\bibliography{bibliography}

\begin{thebibliography}{10}

\bibitem{DeAmortizedCuckooHashingProvableWorstCasePerformanceAndExperimentalResults}
Yuriy Arbitman, Moni Naor, and Gil Segev.
\newblock De-amortized cuckoo hashing: Provable worst-case performance and
  experimental results.
\newblock In {\em International Conference on Automata, Languages and
  Programming (ICALP)}, number 5555 in LNCS, pages 107--118. Springer, 2009.

\bibitem{RobinHoodHashing}
Pedro Celis, Per-Ake Larson, and J.~Ian Munro.
\newblock Robin hood hashing.
\newblock In {\em 26th Symposium on Foundations of Computer Science (FOCS)},
  pages 281--288, Oct 1985.

\bibitem{CuckooHashingFurtherAnalysis}
Luc Devroye and Pat Morin.
\newblock Cuckoo hashing: Further analysis.
\newblock {\em Information Processing Letters}, 86(4):215 -- 219, 2003.

\bibitem{TightThresholdsForCuckooHashingViaXorSat}
Martin Dietzfelbinger, Andreas Goerdt, Michael Mitzenmacher, Andrea Montanari,
  Rasmus Pagh, and Michael Rink.
\newblock Tight thresholds for cuckoo hashing via {XORSAT}.
\newblock In {\em 27th International Conference on Automata, Languages and
  Programming (ICALP)}, pages 213--225, 2010.

\bibitem{DynamicPerfectHashingUpperAndLowerBounds}
Martin Dietzfelbinger, Anna Karlin, Kurt Mehlhorn, Friedhelm~Meyer auf~der
  Heide, Hans Rohnert, and Robert~E. Tarjan.
\newblock Dynamic perfect hashing: Upper and lower bounds.
\newblock {\em SIAM Journal on Computing}, 23(4):738--761, 1994.

\bibitem{CuckooHashingWithPages}
Martin Dietzfelbinger, Michael Mitzenmacher, and Michael Rink.
\newblock Cuckoo hashing with pages.
\newblock In {\em 19th European Symposium on Algorithms (ESA)}, number 6942 in
  LNCS, pages 615--627. Springer, 2011.

\bibitem{BalancedAllocationAndDictionariesWithTightlyPackedConstantSizeBin}
Martin Dietzfelbinger and Christoph Weidling.
\newblock Balanced allocation and dictionaries with tightly packed constant
  size bins.
\newblock {\em Theoretical Computer Science}, 380(1):47--68, 2007.

\bibitem{SpaceEfficientWithWorstCaseConstantAccess}
Dimitris Fotakis, Rasmus Pagh, Peter Sanders, and Paul Spirakis.
\newblock Space efficient hash tables with worst case constant access time.
\newblock {\em Theory of Computing Systems}, 38(2):229--248, 2005.

\bibitem{OnTheInsertionTimeOfCuckooHashing}
Nikolaos Fountoulakis, Konstantinos Panagiotou, and Angelika Steger.
\newblock On the insertion time of cuckoo hashing.
\newblock {\em SIAM Journal on Computing}, 42(6):2156--2181, 2013.

\bibitem{AnAnalysisOfRandomWalkCuckooHashing}
Alan Frieze, Páll Melsted, and Michael Mitzenmacher.
\newblock An analysis of random-walk cuckoo hashing.
\newblock {\em SIAM Journal on Computing}, 40(2):291--308, 2011.

\bibitem{TheAnalysisOfDoubleHashing}
Leo~J. Guibas and Endre Szemeredi.
\newblock The analysis of double hashing.
\newblock {\em Journal of Computer and System Sciences}, 16(2):226--274, 1978.

\bibitem{LessHashingSamePerformance}
Adam Kirsch and Michael Mitzenmacher.
\newblock Less hashing, same performance: Building a better bloom filter.
\newblock In Yossi Azar and Thomas Erlebach, editors, {\em 14th European
  Symposium on Algorithms (ESA)}, number 4168 in LNCS, pages 456--467.
  Springer, 2006.

\bibitem{UsingAQueueToDeAmortizeCuckooHashingInHardware}
Adam Kirsch and Michael Mitzenmacher.
\newblock Using a queue to de-amortize cuckoo hashing in hardware.
\newblock In {\em 45th Annual Allerton Conference on Communication, Control,
  and Computing}, volume~75, 2007.

\bibitem{Knuth3}
Donald~E. Knuth.
\newblock {\em The Art of Computer Programming, Volume 3: (2nd Ed.) Sorting and
  Searching}.
\newblock Addison Wesley Longman Publishing Co., Inc., Redwood City, CA, USA,
  1998.

\bibitem{AlgorithmicImprovementsForFastConcurrentCuckooHashing}
Xiaozhou Li, David~G. Andersen, Michael Kaminsky, and Michael~J. Freedman.
\newblock Algorithmic improvements for fast concurrent cuckoo hashing.
\newblock In {\em 9th European Conference on Computer Systems}, EuroSys '14,
  pages 27:1--27:14. ACM, 2014.

\bibitem{ThePowerOfTwoChoicesInRandomizedLoadBalancing}
M.~Mitzenmacher.
\newblock The power of two choices in randomized load balancing.
\newblock {\em IEEE Transactions on Parallel and Distributed Systems},
  12(10):1094--1104, Oct 2001.

\bibitem{SomeOpenQuestions}
Michael Mitzenmacher.
\newblock Some open questions related to cuckoo hashing.
\newblock In Amos Fiat and Peter Sanders, editors, {\em 17th European Symposium
  on Algorithms (ESA)}, volume 5757 of {\em LNCS}, pages 1--10. Springer, 2009.

\bibitem{LockFreeCuckooHashing}
N.~Nguyen and P.~Tsigas.
\newblock Lock-free cuckoo hashing.
\newblock In {\em 2014 IEEE 34th International Conference on Distributed
  Computing Systems (ICDCS)}, pages 627--636, June 2014.

\bibitem{CuckooHashing}
Rasmus Pagh and Flemming~Friche Rodler.
\newblock Cuckoo hashing.
\newblock {\em Journal of Algorithms}, 51(2):122--144, 2004.

\end{thebibliography}

\appendix
\section{Parameterization of DySECT}
\label{app:param}
In this section, we show additional measurements, using the experiment
described in Section~\ref{sec:exp_ti}. We insert 20\,000\,000 elements
into a previously empty table -- using a dynamic table size.  We show
the measurements for different parameterizations of DySECT and Cuckoo
Tables.  First we show different combinations for $B$ and $H$, here we
also show their find performance. Then we show different displacement
techniques using ($B=8$ and $H=3$).  For random walk displacements we
test an optimistic, and a pessimistic variant.

\begin{figure*}[ht]
  \centering
  \includegraphics[width=0.49\textwidth]{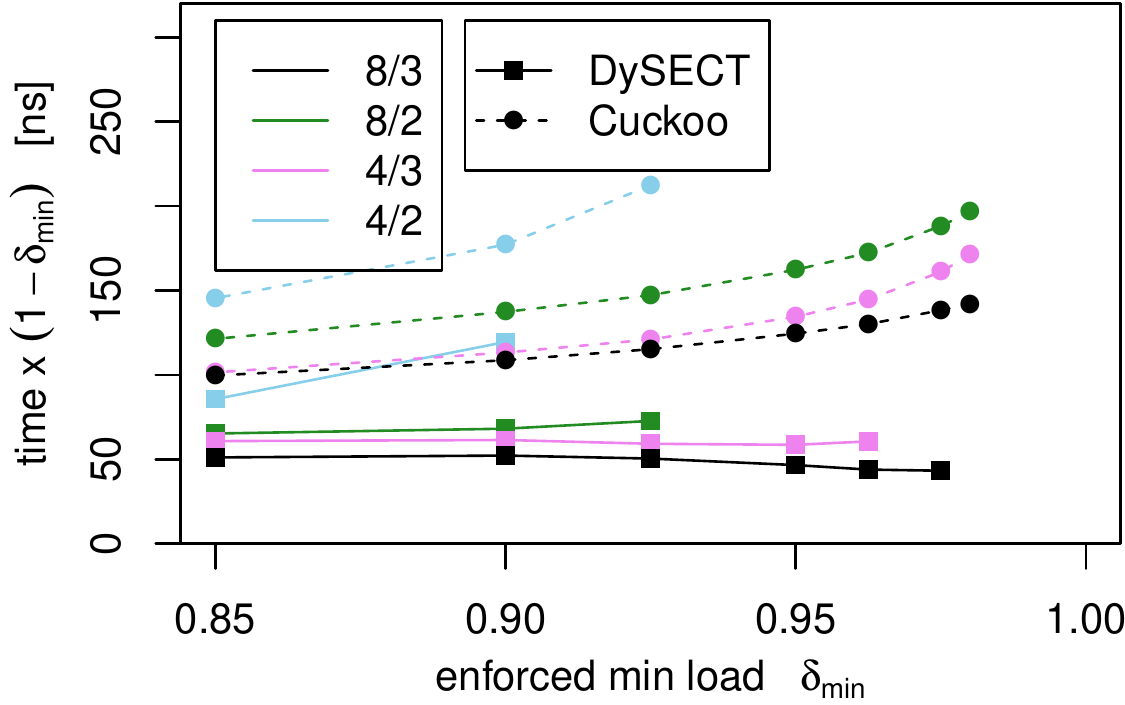}\ \
  \includegraphics[width=0.49\textwidth]{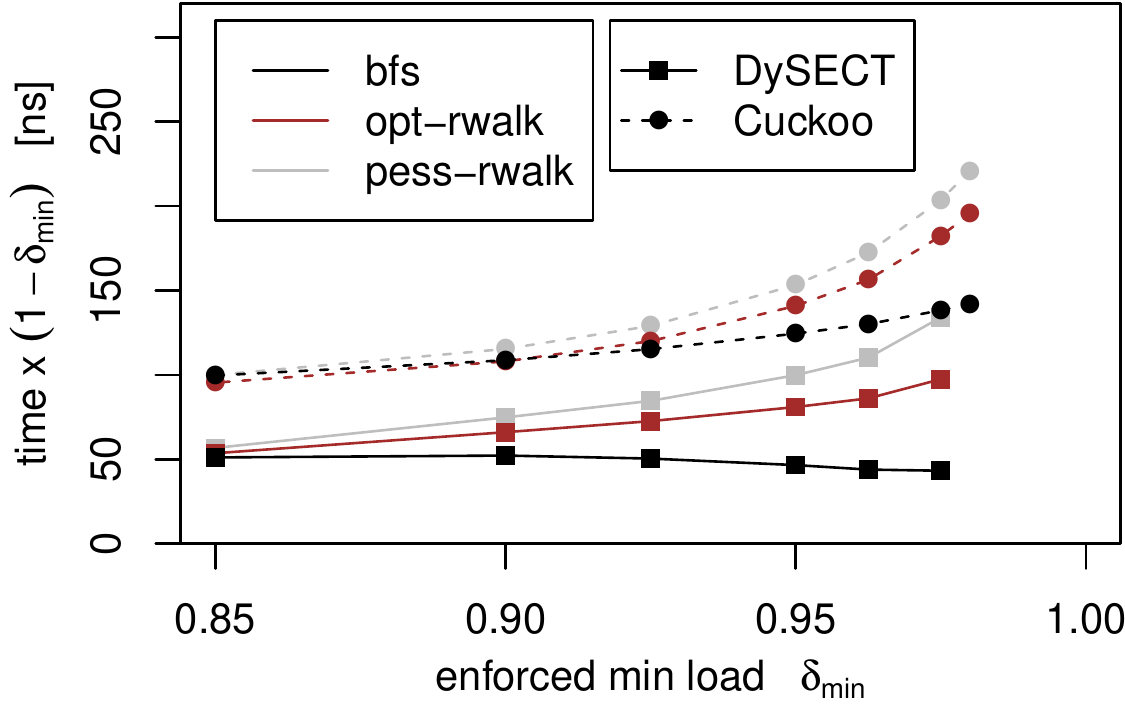}
  \caption{\label{fig:app_insert} Insertions with different $B/H$ Parameterizations (\emph{left}), and different Displacement techniques (using 8/3\emph{right}).  For each displacement, we perform up to 1024 probes for an empty bucket. Measurements end when errors occur.}
\end{figure*}
\begin{figure*}[ht]
  \centering
  \includegraphics[width=0.49\textwidth]{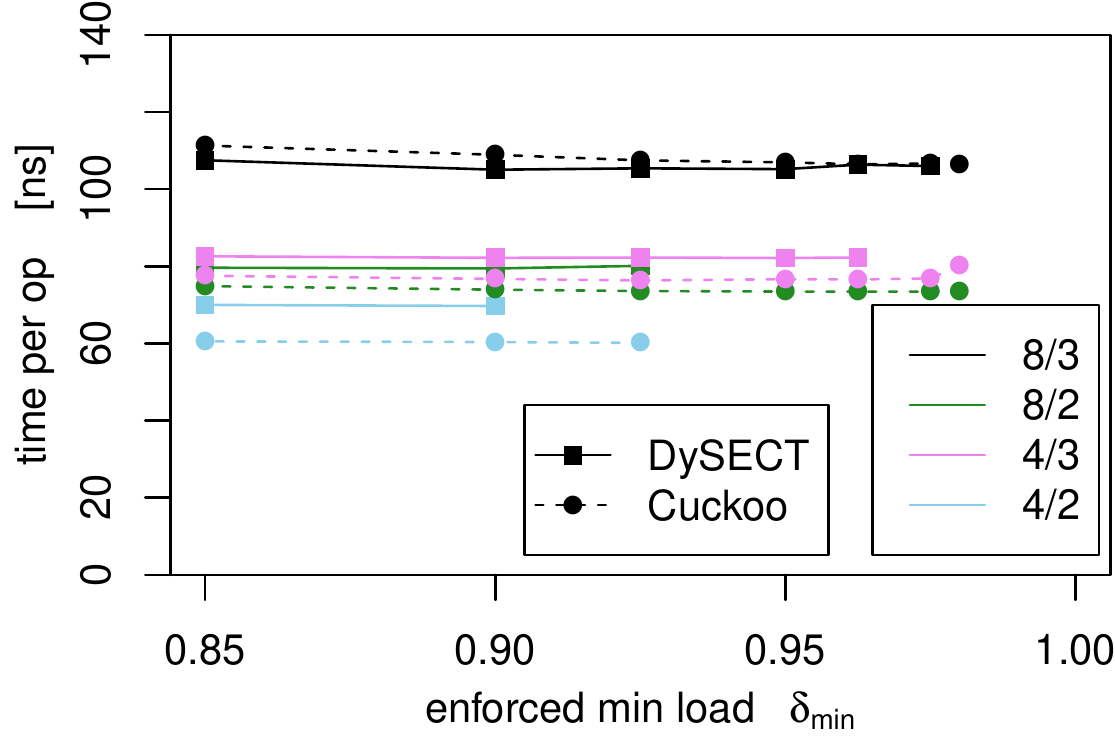}\ \
  \includegraphics[width=0.49\textwidth]{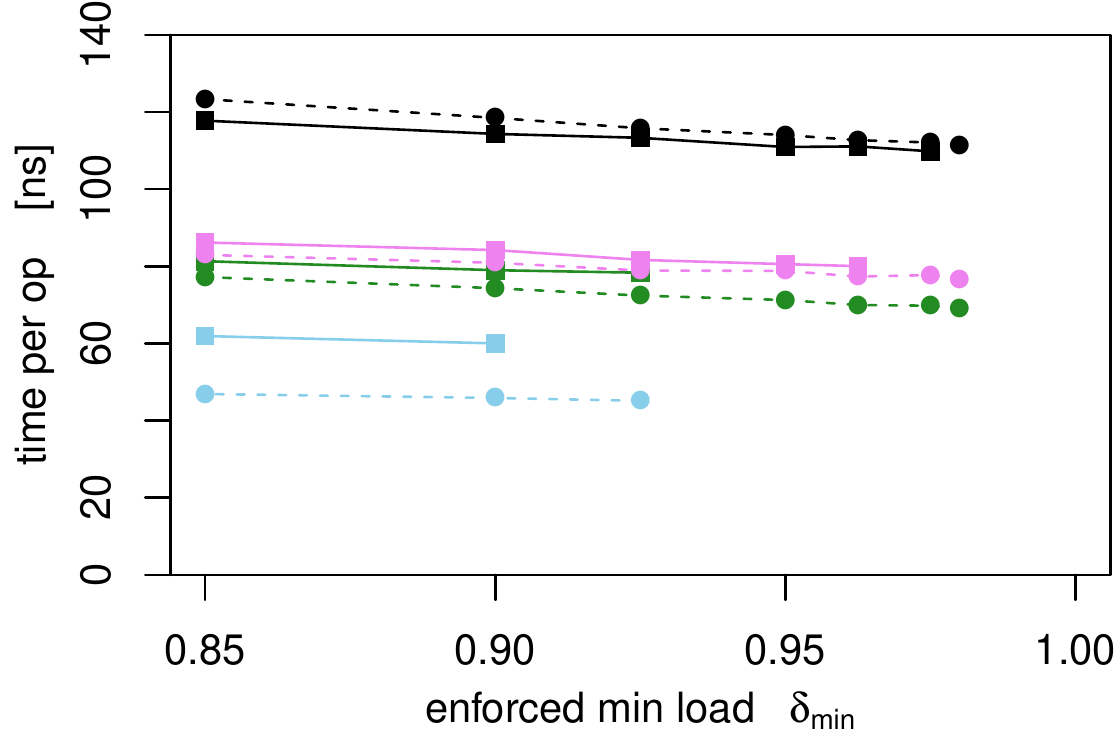}
  \caption{\label{fig:app_find} Finds with different $B/H$ Parameterizations (successful \emph{left}, unsuccessful\emph{right}). Solid lines represents DySECT.  Dashed lines represent bucket cuckoo hashing.}
\end{figure*}

The measurements show, that the chosen parameterization in
Section~\ref{sec:dys_imp} has the best maximum fill bounds and the
best insert performance. Both 4/3 and 8/2 might achieve better
performance on sparser tables, with find heavy workloads.

\end{document}